# Observation of magnetic field-induced contraction of fission yeast cells using optical projection microscopy

Xi Yang,[1] A.W.Beckwith[2]

[1]*Fermi National Accelerator Laboratory, P.O. Box 500, Batavia, Illinois 60510-0500*

[2]*Department of Physics, University of Houston, Houston, Texas 77204-5005, USA.*

*77204-5001, USA.*

The charges in live cells interact with or produce electric fields, which results in enormous dielectric responses, flexoelectricity, and related phenomena. Here we report on a contraction of *Schizosaccharomyces pombe* (fission yeast) cells induced by *magnetic* fields, as observed using a phase-sensitive projection imaging technique. Unlike electric fields, magnetic fields only act on moving charges. The observed behavior is therefore quite remarkable, and may result from a contractile Lorentz force acting on diamagnetic screening currents. This would indicate extremely high intracellular charge mobilities. Besides, we observed a large electro-optic response from fission yeast cells.

A hand placed in front of a diverging beam casts a greatly magnified shadow on the far wall. This principle is exploited in projection microscopy, first proposed in 1939 by Morton and Ramberg [1] with their point projector electron microscope. In 1960, Müller [2] achieved extraordinary magnification and atomic scale resolution using field ion shadow microscopy. More recently, the diffraction limit was attained [3,4] in Fresnel projection electron microscopy, which was used to study defects in carbon nanotubes [5]. Since no lenses are needed between the specimen and the phosphor screen or CCD array, the depth of field is essentially unlimited. In the geometric approximation, the magnification is simply $D/d$, where $d$ and $D$ are the distances from the source to the specimen and to the CCD array, respectively. Interference effects become significant as the object dimensions become comparable to the wavelength [6]. However, an opaque object will be clearly recognizable when the source-object distance is sufficiently small, despite the presence of diffraction fringes around the periphery. This is the limit of Fresnel, or near-field diffraction. For large object-source distances and small object sizes, only a diffraction pattern remains, which is Fraunhofer or far-field diffraction. Fraunhofer will dominate over Fresnel diffraction for an opaque object when $d > a^2/\lambda$, where $a$ is the object dimension and $\lambda$ is the wavelength.

Fig. 1 illustrates how one can implement projection microscopy or diffractrometry at optical wavelengths, by placing the specimen near the focal point of a lens focusing a laser beam. In the experiment shown in Fig. 1, a laser beam was focused onto a live cell by an 18-mm focal length lens, yielding a 3°-divergence angle. The boundaries and internal structures of the cell scatter the incident light to form a boundary diffraction pattern, which can be filtered out in reciprocal space, if necessary. An additional interference pattern, which is affected by the cell's internal dielectric properties and geometry, is created by resonances due to multiple internal reflections. The organisms' mobility was restricted by leaving only a thin film of cell suspension coating the glass plate via capillary attraction. A coil was used to generate the magnetic field along the $z$-axis, and a voltage between two contacts 1-mm apart provided the electric field. The setup allowed one to position the specimen in three dimensions so that the image or diffraction pattern from a single organism could be centered within the field of view. This unusual, phase-sensitive technique provides high contrast without the need for contrast agents, which is advantageous for studying the dynamics of live cells. A transparent organism acts as a complex microlens that produces interference fringes, or contours of constant optical path length as in a Pohl interferometer [7]. The image or diffraction pattern is influenced by the size, geometry, and internal refractive index variations within the specimen, and its position relative to the focal point. The use of Green's functions and singular value decomposition has recently been shown [8] to enable 3-D sample reconstruction from the diffraction patterns obtained in projection electron microscopy. This method, or digital in-line holography techniques [9,10], could potentially enable the construction of real-space image sequences, which would offer advantages for studying live-cell dynamics and *in vivo* processes. Confocal and widefield (deconvolution) microscopes are often too slow, because they require sequential acquisition of many planes of focus to build up a 3-D image.

In one experiment, a 24-gauss magnetic field was applied along the $z$-axis, parallel to the laser beam (see Fig. 1), to fission yeast (*Schizosaccharomyces pombe*) cells placed near the focal point. A 1-frame per second image sequence was acquired with a CCD camera to observe the dynamic response of a single cell to the applied field. The resulting Fresnel-like boundary diffraction patterns were observed to contract along radial directions perpendicular to the *z*-axis, as shown in Fig. 2 ((a)-(d)). Moreover, this contractile motion only stopped when the boundary diffraction pattern disappeared, possibly due to breakage of the ~10-nm thick cell membrane. We studied the magnetic field-induced responses of three additional cells, all of which showed similar contractile behavior, as shown in Figs. 2(e) – 2(p).

The magnetic field response of a fifth yeast cell was then studied using conventional light microscopy, as shown in Fig. 3. The geometry of the microscope necessitated the use of a smaller electromagnet. However, the reduced magnetic field (~8 gauss) enabled the observation of contractile behavior that was approximately reversible when the magnetic field was turned on and off. The use of contrast agents, which might interfere with the cell's intrinsic electromagnetic behavior, was avoided. The gray-scale image was digitally enhanced to compensate for the reduced contrast and depth of field. The apparent cross-sectional area of the organism (in number of pixels) was calculated for each image. This area decreased to 82% of its initial value 4 seconds after the field was applied, and then increased back to 102% of the original size 4 seconds after the field was turned off (see Figures 3 (a) – (f)). Further, systematic studies are needed to determine the relative importance of magnetic field strength and light intensity on the cell responses.

In another experiment, an electric field of $10^4$ V/m was applied along the *x*-axis, perpendicular to the laser beam, and a fission yeast cell was placed near the focal point using the optical projection microscopy setup shown in Fig. 1. In this case the diffraction patterns in Fig. 4, although Fraunhofer-like in appearance, appear to be contained within a well-defined area. This suggests that they result from diffraction by spatial variations in refractive index due to the complex cell structure, rather than being true far-field diffraction patterns of the entire cell. We believe that field-induced forces acting on the charges caused non-uniform stress to buildup within the organism and its membrane. This field-induced stress apparently evolved until attaining equilibrium with the Coulomb forces by redistributing the internal charges, as suggested by Fig. 4(d). Reversing the field direction then appeared to induce birefringence, as shown in Fig. 4(g). Finally, after doubling the electric field strength to $2\times10^4$ V/m, the cells rapidly lost their ability to respond to any electric field. This may stem from cell death and the resulting loss of membrane potential. Similar experiments, to be reported elsewhere, were carried out on *E. coli* bacteria, which exhibited extremely high electro-optic coefficients.

The observed magnetic field-induced contraction is, in our view, the most extraordinary result of the two experiments. Yeast cells are negatively charged and this charge differential between the internal and external environments is maintained by the cell membrane [11]. A rapidly increasing magnetic field would thus induce eddy currents among mobile charges. Any currents flowing parallel to the outer cell membrane, for example, would experience a contractile Lorentz force, which would oppose the increase in magnetic flux through the cell body. What is especially surprising in this case, however, is that the field experienced by the cells in the first experiment initially increased quite slowly, so that $dB/dt$ was small. In addition, most of the apparent contraction occurred after the field had stabilized (when $dB/dt \sim 0$), indicating possible diamagnetic screening currents as in a superconductor. Moreover, the enormous low frequency dielectric constants of fission yeast cell suspensions, which were indicated by the experiment in Fig. 4, are found to be consistent with theoretical predictions [12] only if the intracellular charge mobilities are taken to be extremely high.

Well before the discovery of high-temperature superconductivity, Fröhlich suggested [13-16] that long-range quantum coherence might play a fundamental role in biological systems. This may provide a mechanism to integrate complex interactions that take place within a cell, such as the coordinated action of microtubules. A recent model [17] proposes that microtubules engage in a form of quantum computation. Although somewhat speculative, the plausibility of this hypothesis is supported, at least indirectly, by observations [18-20] that quantum entanglement can be much longer-lived and more robust than previously believed. Further research is clearly warranted to investigate the possible existence of biological diamagnetism.

The authors acknowledge helpful discussions with Bill W. Mayes II and the assistance of Hugo Sanabria in preparing the figures. This work received support from the Robert A. Welch Foundation, the Texas Center for Superconductivity and Advanced Materials, the Texas Higher Education Coordinating Board, the Institute for Space Systems Operations, and the National Institutes of Health.

**Figure Caption:**

FIG. 1. Schematic of the experimental setup for optical projection microscopy and diffractometry. The voltage source and electromagnet were used to study the organisms' responses to electric and magnetic fields.

FIG. 2. A series of Fresnel-like boundary diffraction images, showing the effects of a magnetic field on fission yeast cells. The time at which the field at the cell attains 24 Gauss is defined to be $t = 0$. **Cell #1: a)** Before field was applied. **b)** $t = 6$ seconds. **c)** $t = 12$ s. **d)** $t = 18$ s. **Cell #2: e)** Before field was applied. **f)** $t = 120$ s. **g)** $t = 240$ s. **h)** $t = 360$ s. **Cell #3: i)** Before field was applied. **j)** $t = 20$ s. **k)** $t = 40$ s. **l)** $t = 60$ s. **Cell #4 m)** Before field was applied. **n)** $t = 80$ s. **o)** $t = 160$ s. **p)** $t = 240$ s.

FIG. 3. Conventional light microscopy images of **Cell #5**, in which the contrast was digitally enhanced. **a)** Before magnetic field was applied. **b)** $t = 2$ s. **c)** $t = 4$ s. **d)** Magnetic field turned off. **e)** 2 s after d). **f)** 4 s after d). Calculations of the apparent cross-sectional area of the cell image showed a decrease to 82% of its original size, and a subsequent increase back to 102% of the original area in Fig. 3(f).

FIG. 4. The transmitted diffraction patterns of a fission yeast cell. **a)** Before the electric field was applied. **b)** 60 s after applying a field of $10^4$ V/m to the yeast cell. **c)** 140 s after field was applied. **d)** 290 s after field was applied. **e)** After **d**, switching off the electric field and waiting for 360 s. **f)** After **e**, applying a $10^4$ V/m field in the reverse direction and maintaining the same magnitude for 200 s. **g)** After **f**, reversing the electric field and doubling the field strength for 30 s. **h)** 70 s after **f**. **i)** 230 s after **f**.

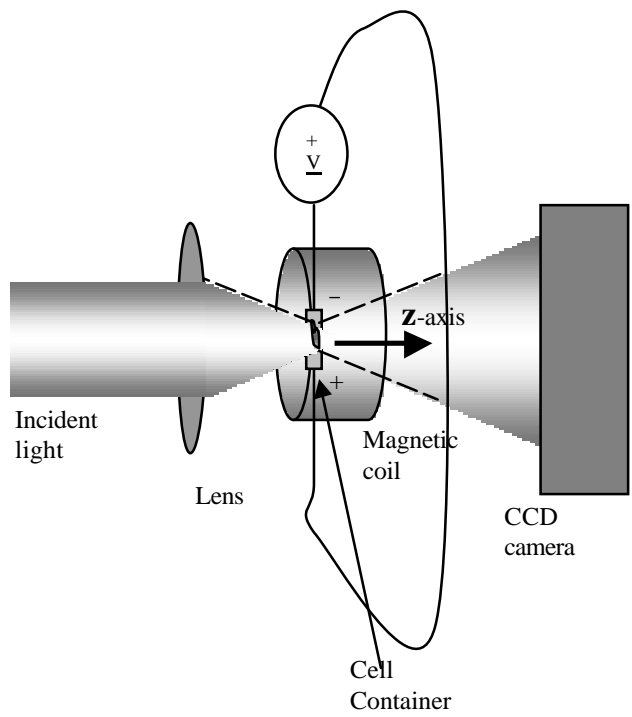

Figure 1

Yang, et al.

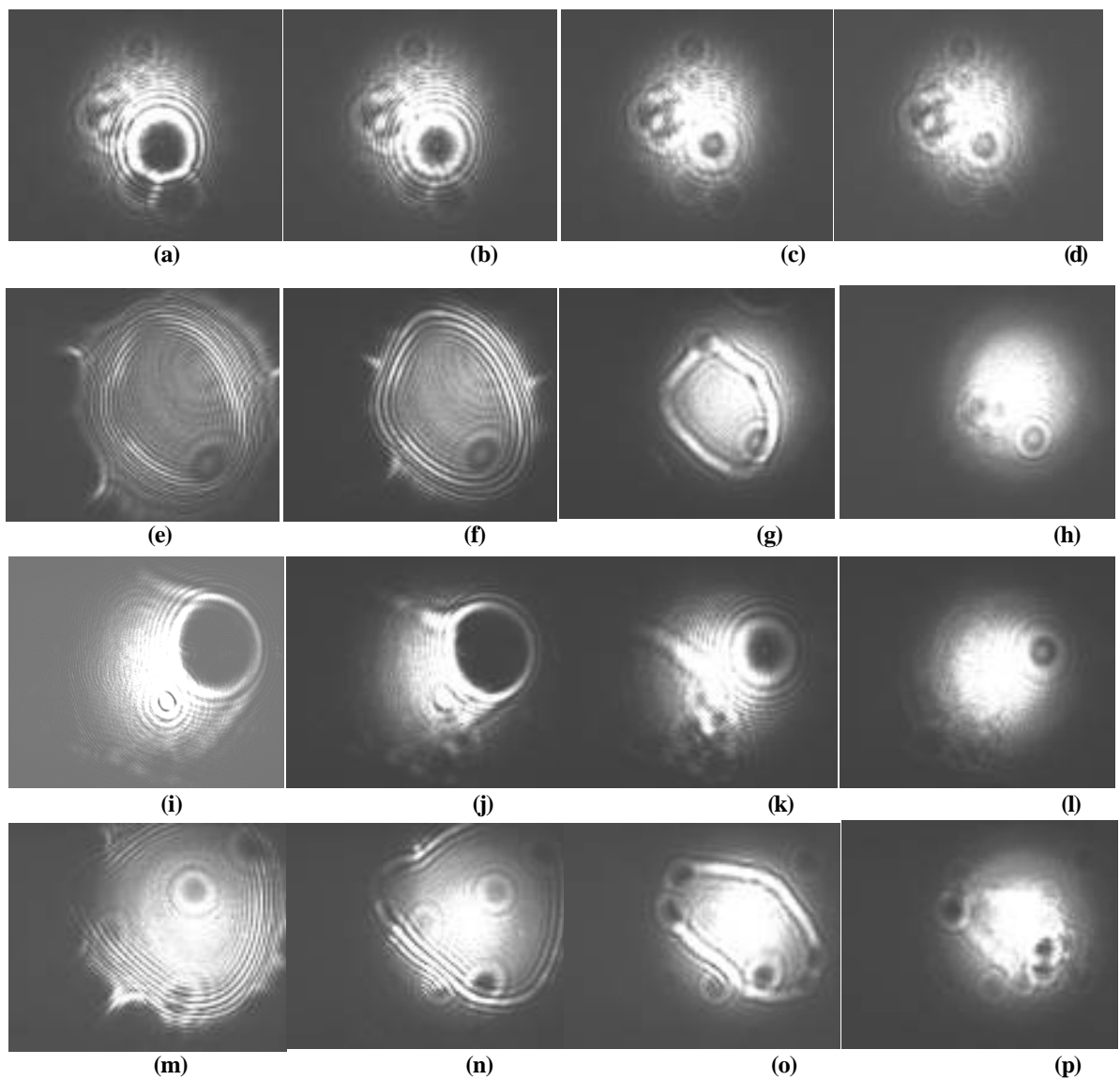

Fig. 2

Yang, et al.

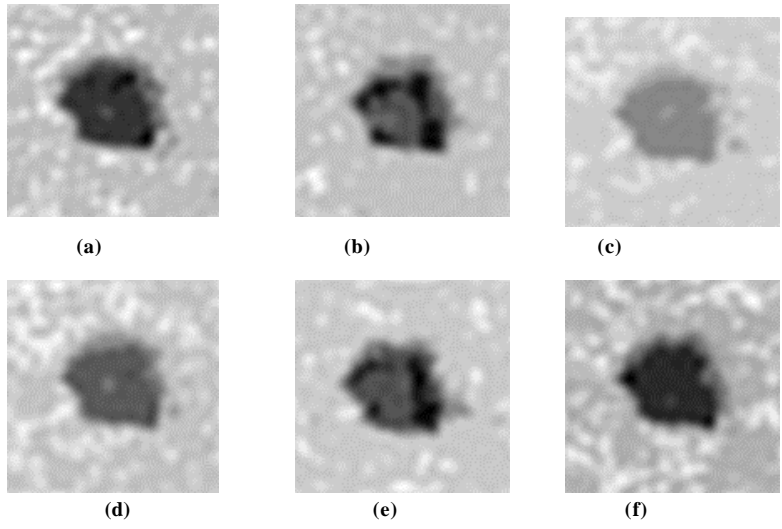

Figure 3

Yang, et al.

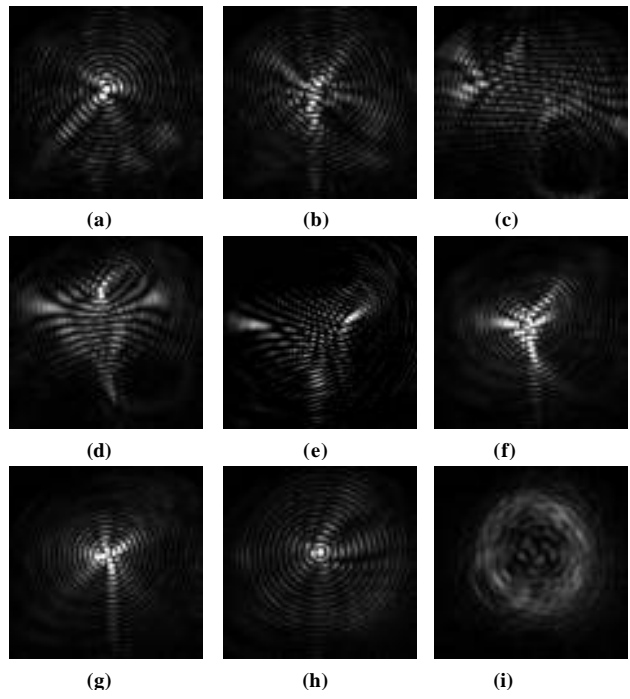

Fig. 4

Yang, et al.